\documentclass[letterpaper,twocolumn,10pt]{article}
\usepackage{usenix-2020-09}

\usepackage{tikz}
\usepackage{amsmath}
\usepackage{cleveref}

\usepackage{graphicx}
\setkeys{Gin}{keepaspectratio}
\usepackage{subfig}

\begin{document}

\date{}

\title{\Large \bf Equilibrium: Optimization of Ceph Cluster Storage by Size-Aware Shard Balancing}

\author{
  {\rm Jonas Jelten} \\
  Technical University of Munich, Germany
  \and
  {\rm Alessandro Wollek} \\
  Technical University of Munich, Germany
  \and
  {\rm David Frank}\\
  Technical University of Munich, Germany
  \and
  {\rm Tobias Lasser} \\
  Technical University of Munich, Germany
}

\maketitle

\begin{abstract}
  Worldwide, storage demands and costs are increasing.
  As a consequence of fault tolerance, storage device heterogenity, and data center specific constraints, optimal storage capacity utilization cannot be achieved with the integrated balancing algorithm of the distributed storage cluster system Ceph.
  This work presents \textit{Equilibrium}, a device utilization size-aware shard balancing algorithm.
  With extensive experiments we demonstrate that our proposed algorithm balances near optimally on real-world clusters with strong available storage capacity improvements while reducing the amount of needed data movement.
\end{abstract}

\begin{figure*}
	\centering
        \includegraphics[width=0.8\linewidth]{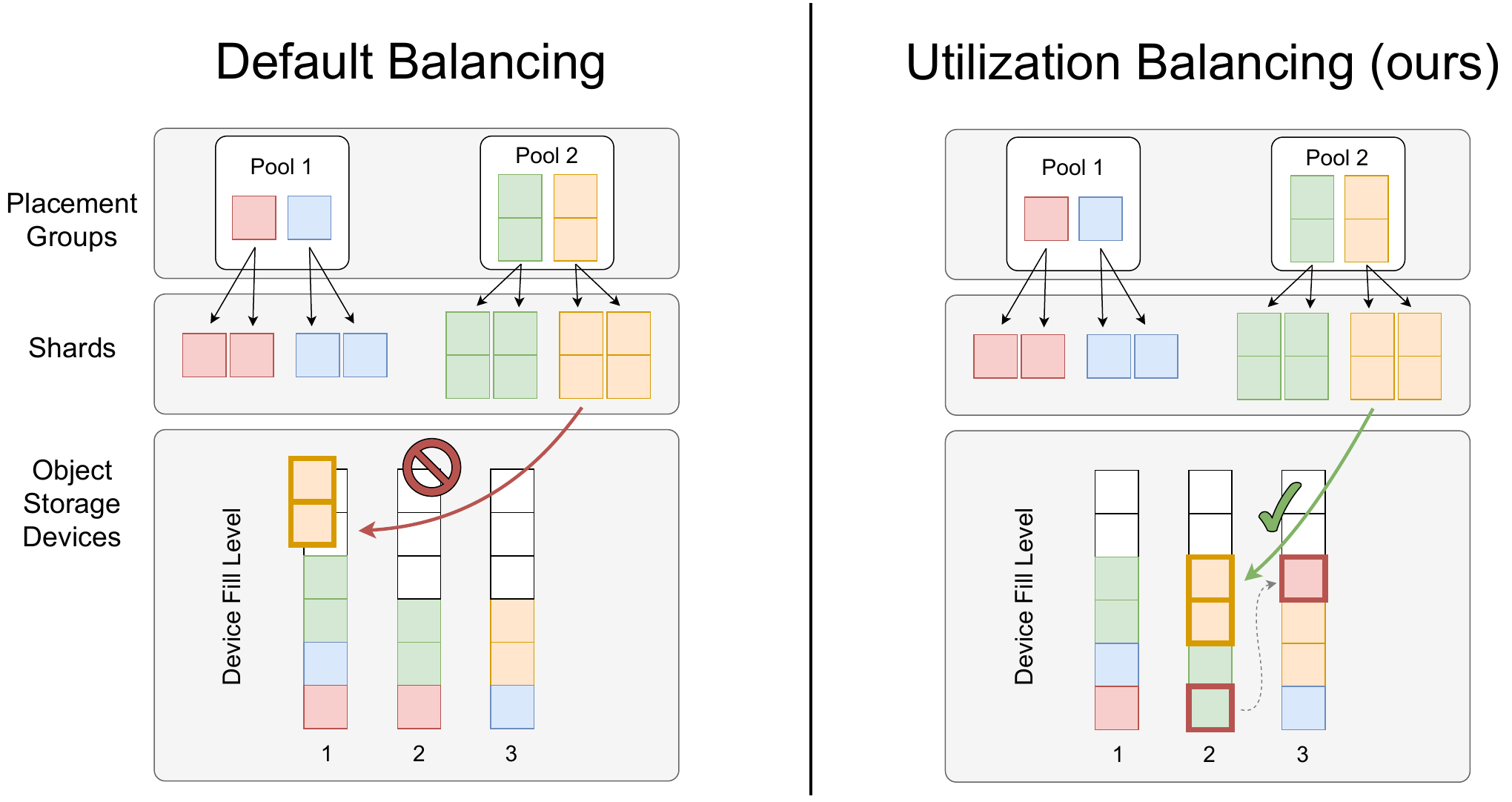}
    \caption{\label{fig:fig01movementidea}
      The default balancer (left) aims to distribute placement groups (PGs) equally without considering the devices' fill level.
      This reduces cluster free space due to uneven device utilization.
      Our proposed balancing algorithm \textit{Equilibrium} (right) additionally includes device fill levels to address this issue.
      This effectively increases the available storage space.
    }
\end{figure*}

\section{Introduction}
With ever increasing storage demands~\cite{patrizio2018idc,kaisler2013big,van2014building}, effective utilization of storage devices and management of storage servers and clusters becomes more important.
The efficient use of existing storage devices, such as hard disk drives, not only reduces per byte storage costs, but also reduces the environmental impact on our planet.
Fault tolerance and other constraints enforce that data cannot be placed on arbitrary storage devices but must be distributed.
Suboptimal distribution fills the storage devices unequally.
This leads to some devices being full, see Figure~\ref{fig:fig01movementidea}.
Hence, one lever for increasing the efficacy of existing devices is an intelligent distribution of data across storage devices.

One leading open-source, distributed storage system designed to provide scalable storage is \href{https://ceph.com/}{Ceph}~\cite{weil2006ceph}. Ceph provides a decoupling of the data from the physical storage hardware, allowing for a high degree of fault tolerance and high availability of the storage. It provides interfaces for multiple storage solutions. Ceph exhibits robust scalability, capable of handling increasingly larger clusters without compromising its performance.

Ceph's standout feature lies in its versatility, offering support for object, block, and file storage in a single unified platform, making it adaptable to diverse storage requirements. As an open-source solution, Ceph stands out for its cost-effectiveness, allowing organizations to deploy scalable storage without the licensing costs associated with proprietary and commercial alternatives. Ceph's distributed architecture and data redundancy mechanisms contribute to its exceptional scalability and fault tolerance, ensuring data integrity and high availability.

Ceph has the built-in \texttt{mgr balancer}, which provides basic functionality to deal with the unequal distribution of data across storage devices. However, the algorithm is based on an ideal setting that does not hold to real world data clusters, leading to suboptimal storage availability and increased storage costs.

In this work, we propose \textit{Equilibrium}, a novel data balancer that optimally distributes across storage devices, increasing available storage on existing systems.

This work makes the following contributions:
\begin{itemize}
  \item We present \textit{Equilibrium}, an effective balancing algorithm designed for Ceph.
  \textit{Equilibrium} optimizes cluster storage efficiency by iteratively redistributing shards from heavily utilized to underutilized Object Storage Devices (OSD), all while adhering to the constraints defined by the CRUSH rules.
  \item We demonstrate that our proposed balancer outperforms the native \texttt{mgr balancer} by strongly increasing available free storage space and minimizing OSD utilization variations across the entire cluster.
  Unlike the built-in \texttt{mgr balancer}, \textit{Equilibrium} achieves balancing with explicit shard movements.
  \item \textit{Equilibrium} is already being used by multiple data centers addressing their critical need for efficient OSD utilization, making it a valuable contribution to any Ceph user.
\end{itemize}

\section{Background}
\subsection{Ceph}
Ceph is a distributed storage cluster system, where data is stored on Object Storage Devices (OSDs) across many servers~\cite{weil2007rados}.
Physically, OSDs are, for example, solid state drives (SSDs) or hard disk drives (HDDs).
Ceph allows clients to access four kinds of storage services:
\begin{itemize}
    \setlength{\itemsep}{0pt}
    \item Block device (\href{https://docs.ceph.com/en/latest/rbd/}{RBD}),
    \item Network filesystem (\href{https://docs.ceph.com/en/latest/cephfs/}{CephFS}),
    \item Object gateway (\href{https://docs.ceph.com/en/latest/radosgw/}{RGW, S3, Swift}),
    \item Raw key-value storage via \href{https://docs.ceph.com/en/latest/man/8/rados/}{(lib)rados}.
\end{itemize}
The common ground of these services is that all store their data as ``objects'', usually 4 MiB in size.
Larger blocks or files are split up into several smaller objects.

From a high-level perspective, Ceph can be described as a distributed key-value database.
Hierarchically, a Ceph cluster is partitioned into pools, placement groups (PGs), and shards (see Figure~\ref{fig:fig01movementidea}).
Pools have custom redundancy rules and access control, and they can contain namespaces for stored objects.

Objects are stored by Ceph in a scalable and failure resistant manner.
Within a pool, objects are placed in PGs by hashing the object's key (``name'') and taking the last $x$ bits of the hash to choose the placement group (hence $2^x$ PGs).
More PGs mean better distribution, but they also mean processing overhead on OSDs.
Each PG resides on multiple OSDs through its shards, they are placed depending on the pools redundancy rules: either by copy or with erasure coding.

Since all stored data is chunked into objects with unique names, and PGs are chosen randomly by the object's name, each object is assigned to a different PG and therefore set of PG shards.
Thus, a client accessing an object communicates with different OSDs.
That way, all requests are spread across all OSDs of the whole cluster.

When a single OSD fails, the missing copy can be automatically recreated on another OSD.

The distribution of objects follows user-defined constraints, the so-called ``CRUSH rules''.
For example, a CRUSH rule could specify that there should be three copies on separate servers (3 OSDs) or RAID6 on twelve server racks (12 OSDs).

\begin{figure*}
	\centering
        \includegraphics[width=\linewidth]{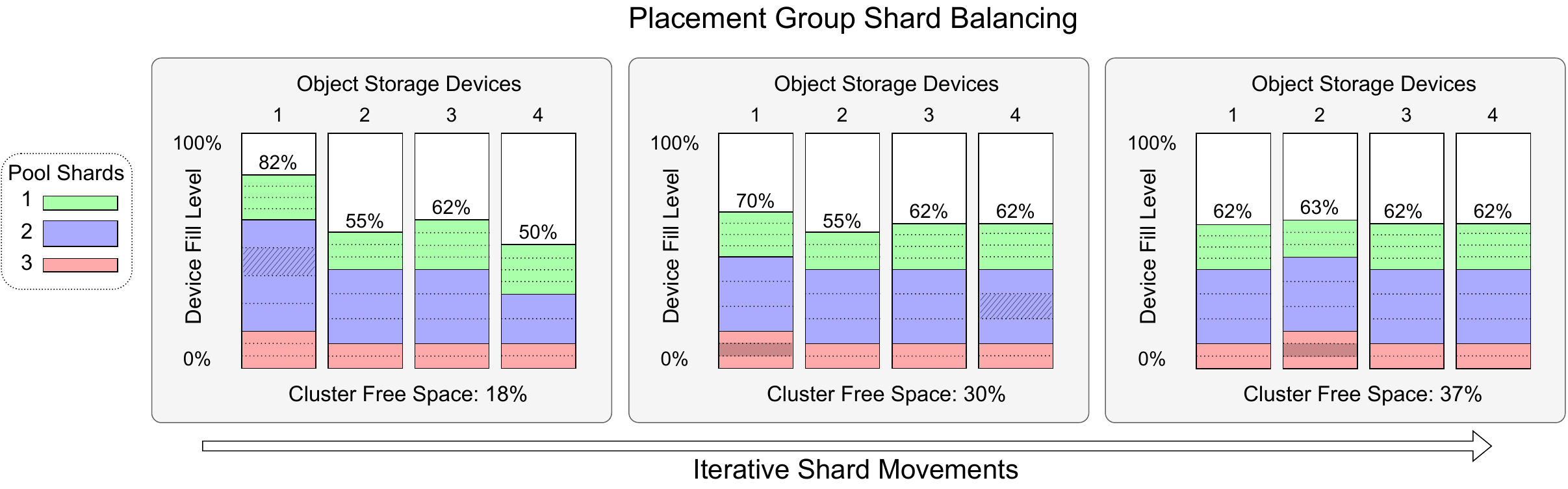}
    \caption{\label{fig:fig02generating_balance}
      Due to probabilistic distribution of newly stored data in placement groups, the fullest object storage device (in this example: OSD 1) limits overall available storage (left).
      Rearranging placement groups to equalize relative device utilization (right) leads to more available storage.
    }
\end{figure*}

Free space in a pool is limited by the most filled OSD having a shard of that pool.
If just a single shard has not enough space to allocate, the pool is full.
Consequently, available storage is maximal when all OSDs are
equally full, see Figure~\ref{fig:fig01movementidea}.
In an existing cluster, storage can be recovered by moving shards across OSDs to reach equilibrium, see Figure~\ref{fig:fig02generating_balance}.

\subsection{The Balancing Problem}

CRUSH is Ceph's data placement algorithm to organize placement groups (PGs) across data centers, racks, servers, and OSDs in a tree structure.
Each node has the weight of the size of all OSDs below it.
PGs are then distributed evenly, weighted by the node size at each tree level.
That way large data centers get more data than small servers, but the relative amount is the same.
Hence, in theory, each OSD should be filled exactly the same relative amount, for example, all are 30~\% full.

To evenly distribute load and utilization, each OSD has an ideal PG shard count for a pool.
This ideal shard count can be calculated with the number of shards in the pool, multiplied with the fraction of the OSD's size and the size of all OSDs in the pool (\texttt{osd\_ideal\_shard\_count} = \texttt{pool\_shard\_count} $\times$ (\texttt{osd\_size} / sum(\texttt{all\_osd\_sizes}))).

In practice, OSDs are not filled equally.
The reasons are manifold:
Between pools, the shards can differ in size and shards cannot be distributed perfectly equal across OSDs.
Additionally, fault tolerance and CRUSH rules can prevent placement in a more equal distribution.
Finally, OSDs can physically differ in storage capacity, while PG shard sizes in a pool are almost equal.

Since the available pool storage is limited by the fullest participating OSD, it is desirable to have no fullest OSD, or in other words, have them all balanced equally.

Since the base CRUSH distribution doesn't guarantee equal data distribution~\cite{weil2006crush}, especially when pools grow and shrink independently, a separate balancing algorithm has to be applied to improve cluster balance.

\subsection{Balancing Algorithms}

Balancing algorithms adjust data placement in a Ceph cluster, according to their specific optimization goal.
Most common goals are capacity maximization, latency minimization, or throughput maximization.
Since the atomic movement unit is a PG shard, balancers can only move these to other locations.
At the same time it is important to not violate any CRUSH rules.
In the ideal case, the balancing algorithm will reach a perfect distribution according to its goal, but as discussed previously this might not be possible in all cases.
Then, the ideal should be approached as closely as possible.

The need for improving balance of capacity in a cluster was noticed early, so Ceph ships with an integrated balancer running in its \texttt{mgr} component.

\subsubsection{Ceph's ``mgr balancer''}

As a mitigation for non-ideal placement through random assignment with CRUSH, Ceph has a built-in tool (\texttt{mgr balancer}), which optimizes by PG shard count on OSDs.
It does so by operating on each pool independently, where it tries to move PGs between OSDs in such a way that the ideal PG shard count is approached.
In theory, this counters random non-ideal assignments by CRUSH.

\begin{figure*}[!ht]
    \centering
    \includegraphics[width=\linewidth]{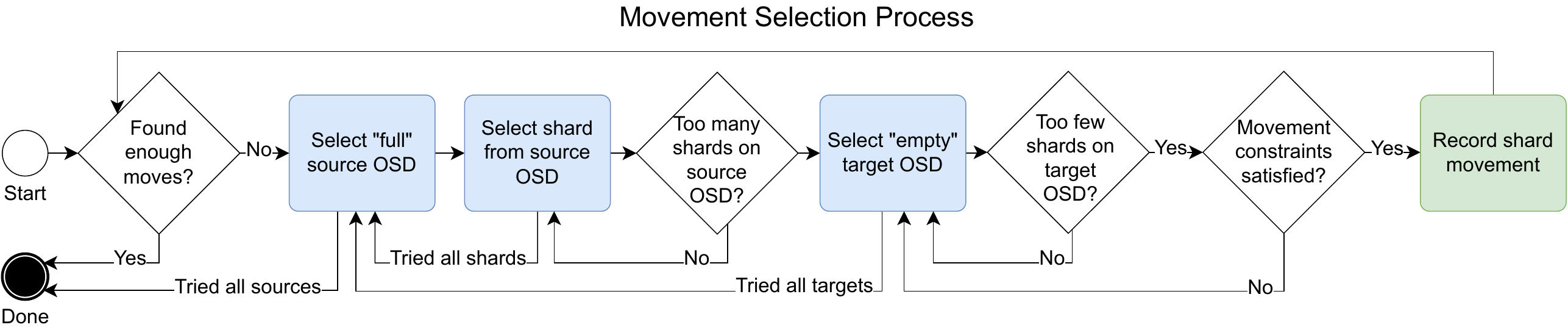}
    \caption{\label{fig:fig03iterative_movements}
      To perform each data movement, an OSD that is filled up the most is selected.
      Then a (preferably large) placement group shard is attempted to be moved to the emptiest possible target OSD.
      A movement is only possible if the number of shards becomes more equally distributed and all data redundancy constraints (the shard's pool CRUSH rule) are respected.
    }
\end{figure*}

As discussed above, perfect distribution according only to the ideal PG shard count is practically impossible.
This is a fundamental limitation of the built-in balancer.
It is important to note, that the built-in tool only has a local storage pool-wide view, i.e. the optimization is performed without checking other pools' distribution. 
At worst, one OSD ends up having larger than ideal PG shard count for each pool in the cluster, even after performing the balancing optimization. 
As available capacity is determined by the most filled OSD, this balancing strategy will limit the free storage of the entire cluster.

Additionally, the \texttt{mgr balancer} does not takes the size of OSDs or PG into consideration.
This is important, specifically, if OSDs are heterogeneous in size or PGs are of very different sizes.
The balancer could move a large PG shard onto a small drive, instead of moving it onto a larger available device.
Though it might optimize the PG shard count of the OSD, it might negatively affect the utilization variance.
A balancer which considers size could have considered a move to a large OSD and optimize both PG shard count and utilization variance.

Finally, there is a candidate selection limitation in \texttt{mgr balancer}: If due to CRUSH rules, sub-trees in the storage hierarchy are little utilized, other sub-trees will not even be considered for balancing, even if selecting one could gain a lot of space. This can happen if you have servers of different sizes in the same pool. Then, due to the defined CRUSH rules (e.g. to uphold a certain redundancy), \texttt{mgr balancer} can not find a valid move for the OSD with the most free space, it will abort instead of looking for other possible devices.

\subsubsection{Other Balancers}


In recent years, several balancing algorithms have been proposed to improve key metrics of a Ceph cluster.
However, to the best of our knowledge, no other work addressed the optimization of available storage capacity by fine-grained control of shard placement.

Flores proposed a balancing strategy that aims for not only write but also balancing read speed from OSDs~\cite{flores2023readbalance}.
In this strategy, the first shard of each PG is well distributed so read accesses are spread across the cluster for optimal performance.
However, their method does not improve the available storage capacity.

Wu et al.~took the real-time network conditions and load status further into consideration when selecting OSDs~\cite{wu2017optimization}.
In another study, Wu et al.~proposed a dynamic data distribution strategy for hybrid object storage systems~\cite{wu2017boss}.
Unlike traditional uniform data distribution methods, this approach dynamically distributes and migrates data objects based on real-time data access patterns.
In contrast to our work, they do not address the reduction in available storage due to uneven data distributions.

Li and Wang proposed to add performance weights based on node heterogeneity, network state, and node load into the CRUSH algorithm to adjust data distribution~\cite{li2023adaptive}. Their optimization goal is maximum I/O performance.
Wang et al.~extended CRUSH by using an extra time-dimension mapping for better data migration control~\cite{wang2020mapx}. They optimize for performance by reducing the data migration load.

Park et al.~utilized a reinforcement learning method to automatically adjust data distribution in response to changing workload patterns in a tiered storage architecture~\cite{park2022reinforcement}.

\section{Method}
\subsection{The Equilibrium Balancer}

The core idea of the \textit{Equilibrium} balancing strategy is to optimize for equal OSD utilization in the cluster, illustrated in Figure~\ref{fig:fig02generating_balance}.
As the storage availability is determined by the fullest OSD, equal distribution maximizes storage availability.
CRUSH distributes the PG shards to OSDs, but the results are not ideal.
Ceph's built-in balancer refines the PG shard distribution, but does not consider the utilization of each device.
Our proposed balancing algorithm, \textit{Equilibrium}, iteratively generates adjustments to the default CRUSH distribution.
The balancer picks suitable PG shards from full OSDs using the iterative scheme described in Figure~\ref{fig:fig03iterative_movements}.
This is a novel approach to find a solution to the balancing problem in Ceph.
Like the built-in balancer, PG shards are moved in such a way that no CRUSH rules are violated and OSDs approach their ideal pool PG shard counts.

\paragraph{Source selection.}
At first, we sort all OSDs based on their relative utilization, expressed as the ratio of used space to device capacity (\texttt{used\_space} / \texttt{device\_size}), within the cluster's target state.
Subsequently, we identify the most heavily utilized OSD as the source candidate.
From this source OSD, we systematically evaluate and select the largest PG shard that can be considered for movement.

\paragraph{Destination Assignment.}
Our balancer determines the destination OSD by adhering to a set of stringent criteria:

\begin{itemize}
    \item Ensuring compliance with the pool's CRUSH rules.
    \item Improving the ideal pool PG shard count for the source and destination OSD.
    \item Enhancing the variance of OSD utilization across the entire cluster.
\end{itemize}

The process involves a sequential evaluation, starting with the largest PG shard on the source OSD.
In cases where relocating the largest shard would violate the specified constraints, we proceed to consider the next largest PG shard.
This iterative selection continues until a suitable shard is found within the $k$-most utilized OSDs.
If not suitable shards are found, the process is terminated.

\paragraph{Shard Movement.}
Once a suitable PG shard for relocation to its designated destination OSD is found, a PG shard move is recorded.
Then the relative cluster utilization is recalculated in line with the generated new cluster state, so that the next move can be generated.
This iterative sequence produces successive shard movement instructions that eventually converge.

The output is a series of movement instructions which trigger the transfer of PG shards from their current OSDs to the designated devices.

This algorithm improves the PG shard count towards to the ideal, while concurrently eliminating OSD utilization disparities across the cluster, thereby optimizing both available storage space in all pools and the distribution of data across available devices to optimize read/write performance.

In scenarios where the $k$-fullest OSDs reach a point where further data evacuation is unfeasible, \textit{Equilibrium}'s runtime complexity is expressed as
\(
\mathcal{O}(k\cdot\texttt{OSDs}\cdot\texttt{PGs}\cdot\texttt{logn}(\texttt{PGs}))
\)
where $k$ is the number of attempts --- essentially denoting the number of source OSDs scrutinized in descending order of capacity utilization. This parameter $k$ can be adjusted to cater to specific termination conditions.


\subsection{Experiments}

To test our balancer's operation, we apply it to several different clusters and evaluate the resulting space and the utilization variance after simulating all data movement.

The clusters have different characteristics:
\begin{itemize}
    \setlength{\itemsep}{0pt}
    \item Cluster A: 225 PGs, 14$\times$HDD 68TiB, 7 pools, 2 with user data
    \item Cluster B: 8731 PGs, 810$\times$HDD 5PiB, 185$\times$SSD 1PiB, 94 pools, 55 with user data, 40 with metadata, 3 with $\sim$1PiB of data
    \item Cluster C: 1249 PGs, 40$\times$HDD 164TiB, 10$\times$NVMe 9TiB, 10 pools, 3 with user data
    \item Cluster D: 4181 PGs, 246$\times$HDD 621TiB, 60$\times$SSD 105TiB, 11 pools, 6 with user data, has hybrid class storage 1SSD 2HDD
    \item Cluster E: 8321 PGs, 608$\times$HDD 8.04PiB, 9$\times$SSD 4TiB, 3 pools, 1 with user data
    \item Cluster F: 577 PGs, 78$\times$HDD 425TiB, 3 pools, 1 with user data
\end{itemize}

\begin{table*}[t]
\centering
\begin{tabular}{|c|r|r|r|r|}
\hline
Cluster       & \multicolumn{2}{c|}{Gained Free Space (TiB)}  & \multicolumn{2}{c|}{Movement Amount (TiB)}  \\
              & Default Method    & Our Method              & Default Method    & Our Method            \\
\hline
A             & 18.2    & \textbf{23.9}                     & \textbf{1.6}   & 1.7                      \\
B             & \textbf{1376.3}  & 925.8                    & 287.1 & \textbf{125.1}                    \\
C             & 8.3     & \textbf{10.4}                     & \textbf{0.8}   & 1.1                      \\
D             & 0.0     & \textbf{4.7}                      & 0.0   & \textbf{0.2}                      \\
E             & 220.8   & \textbf{287.8}                    & 23.7  & \textbf{22.1}                     \\
F             & 65.7    & \textbf{67.5}                     & 13.3  & \textbf{12.8}                     \\
\hline
\end{tabular}
\caption{\label{tab:moveresults} Generated data movement amounts and resulting sum of gained pool space. Our method generally gains more space than the default balancer.
  Although for Cluster B the default balancer did generate more space for all pools overall, it did so mostly for metadata pools with few placement groups and not for ``big'' pools with the most placement groups, where gaining more space would be most relevant (see Figure~\ref{fig:clusterB}).
Better values are highlighted in bold.
}
\end{table*}

Test movement generation with Ceph's default balancer was generated from Ceph repo version v17.2.6 compiled with GCC v12.2.1.
Standalone invocation of the \texttt{mgr balancer}:

\texttt{osdmaptool <testosdmap> --upmap <movement\_result> --upmap-max 10000 --upmap-deviation 1}

This invocation optimizes to ideal pg counts (with a maximum offset of 1).
For all tests, less than 10,000 movements were generated.

Our balancer is implemented in Python v3.10.12 and was configured to terminate after the $k=25$ fullest OSDs could no longer be emptied.

Both balancers start with the same cluster state and terminate once they do not find any more optimization steps.
After movement instructions were generated, their effects were applied in a simulated Ceph cluster in order to measure the movement amount, to predict the resulting free space, and to track OSD utilizations and their variance.
This simulation uses the same cluster state as passed to the balancers.

\section{Results}

The results for all our test clusters are shown in Table~\ref{tab:moveresults}.
For all clusters, except cluster B, our proposed method increased the available pool free space compared to the default balancing method.
These results were achieved while requiring less or a similar amount of data to be moved.

\subsection{Cluster A}
\begin{figure*}
    \centering
        \includegraphics[width=0.49\linewidth]{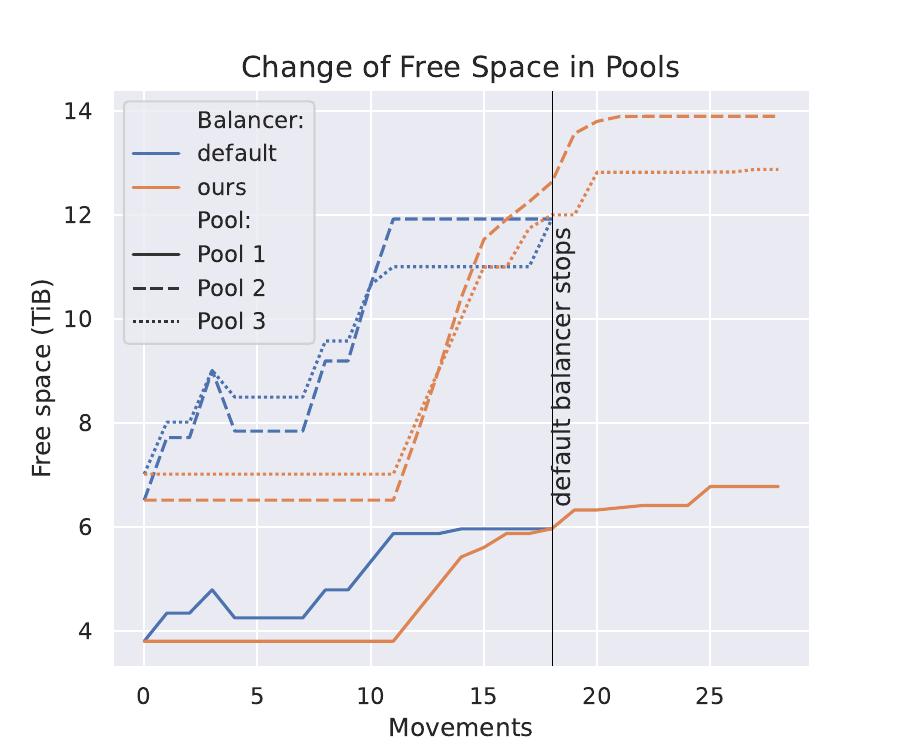}
        \vspace{1ex}
        \includegraphics[width=0.49\linewidth]{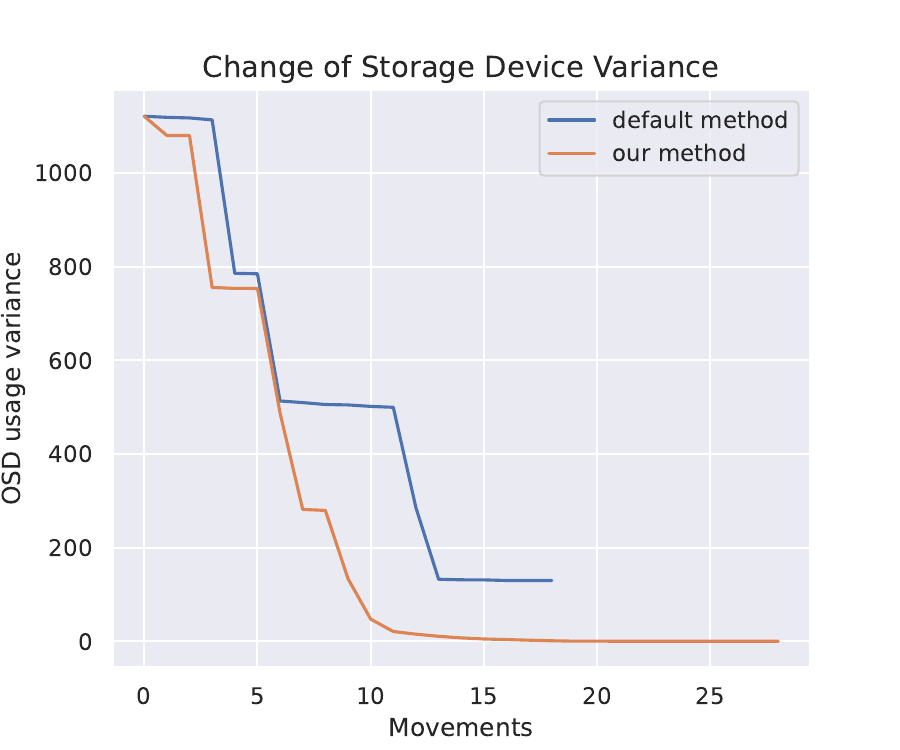}
    \caption{\label{fig:clusterA} During optimizatioin of \textbf{cluster A}, the default balancer stops after 18 moves, but our algorithm continues to find improvements and thus unlocks more available storage in all 3 pools (left). This is confirmed by a lower utilization variance through our balancer (right). }
\end{figure*}

Figure~\ref{fig:clusterA} (left) shows the change in free space per pool depending on the balancing algorithm on cluster A.
Both methods increase the available space by suggesting subsequent movements.
In contrast to the default \texttt{mgr balancer}, our proposed \textit{Equilibrium} balancer continues with optimizing shard distribution after 18 moves.
Comparing the final available space, \textit{Equilibrium} freed up more space across all pools with a difference of up to 16 \%.

Figure~\ref{fig:clusterA} (right) compares the change of OSD variance of cluster A across movements.
While both methods reduce the inequality (variance) of OSD utilization, our proposed method achieved a near perfect balancing variance close to zero.

\subsection{Cluster B}
\begin{figure*}
    \centering
        \includegraphics[width=0.49\linewidth]{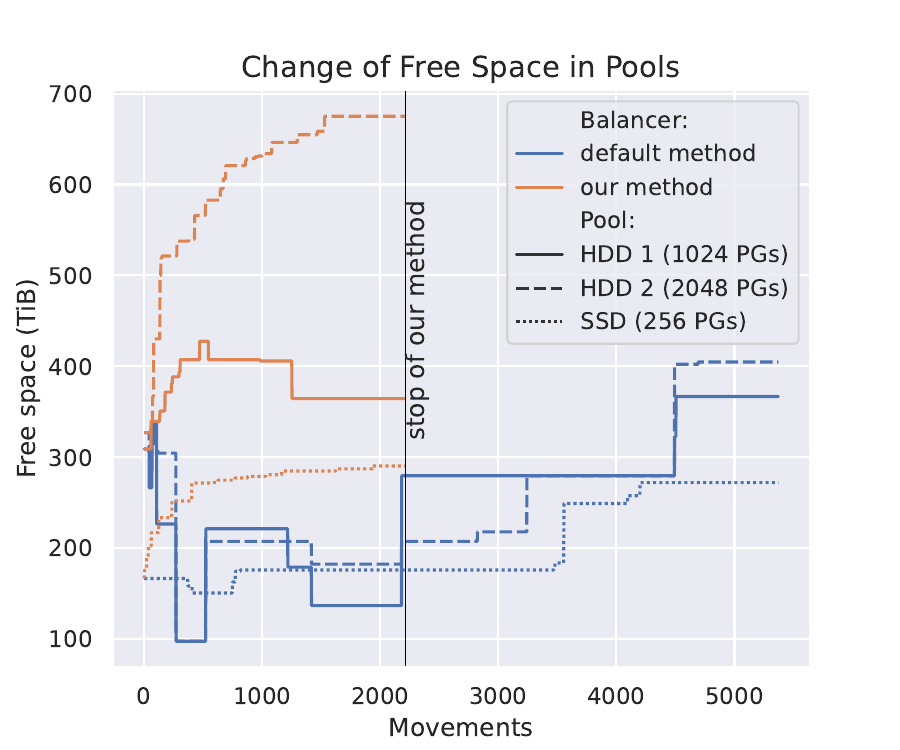}
        \vspace{1ex}
        \includegraphics[width=0.49\linewidth]{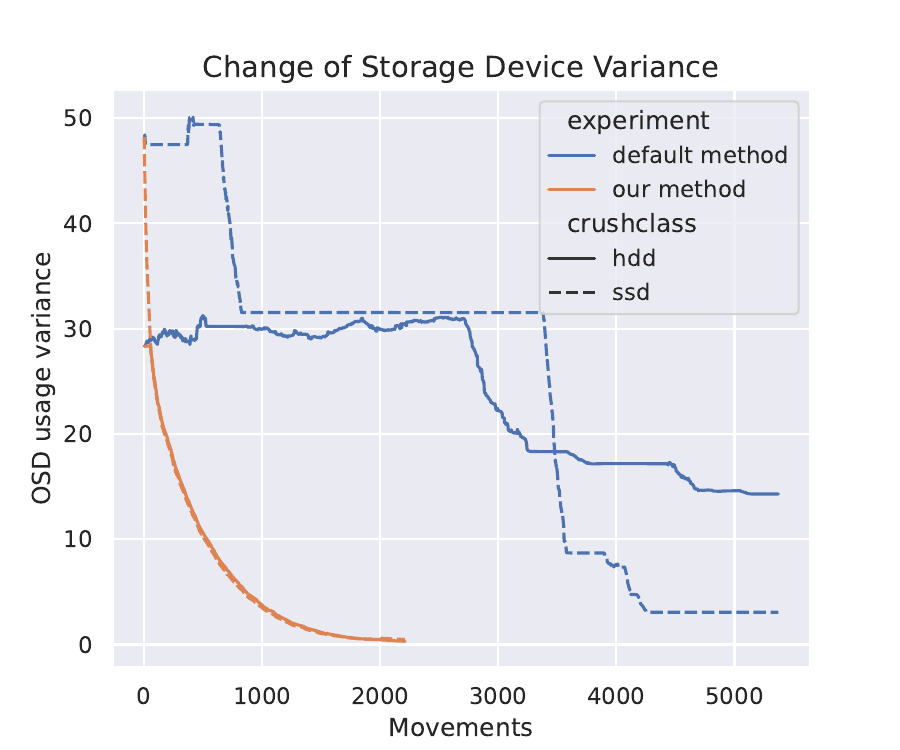}
        \caption{\label{fig:clusterB} Our balancer stops earlier than the default balancer when optimizing \textbf{cluster B}, but achieves more unlocked storage (left) and lower utilization variance (right) earlier.
          Our method also optimizes both SSD and HDD storage utilization simultaneously.
          All smaller pools with 256 or less placement groups are not shown to increase readability.}
\end{figure*}

The effect of balancing algorithms on free space and OSD variance on cluster B are shown in Figure~\ref{fig:clusterB}.
For clarity, small pools with 256 PGs or fewer are not shown.
Using less than half the number of movements, balancing with \textit{Equilibrium} resulted in higher free space with an increase of more than 50 \% compared to the default balancer.
Further evidence is provided by the drop in OSD variance close to zero just after 2000 movements.
In contrast, \texttt{mgr balancer} did not achieve a zero variance even after over 5000 movements.
Furthermore, \textit{Equilibrium} optimized both HDD and SSD utilization simultaneously.

\subsection{Movement Calculation Time}
\begin{figure*}
    \centering
    \subfloat[Cluster A]{
        \includegraphics[width=0.49\linewidth]{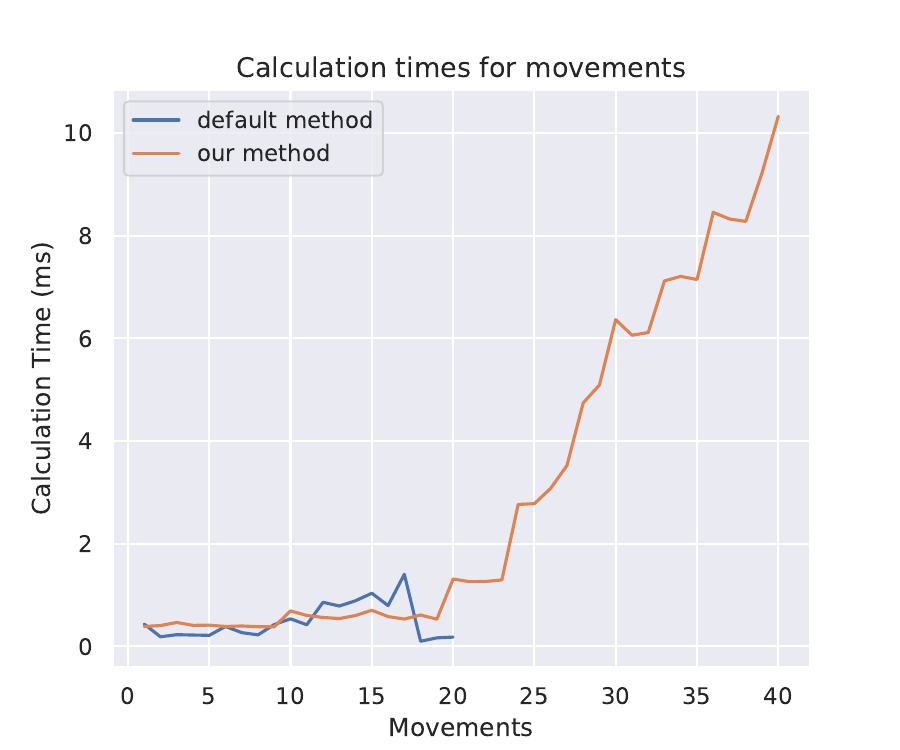}} \vspace{2ex}
    \subfloat[Cluster B]{
      \includegraphics[width=0.49\linewidth]{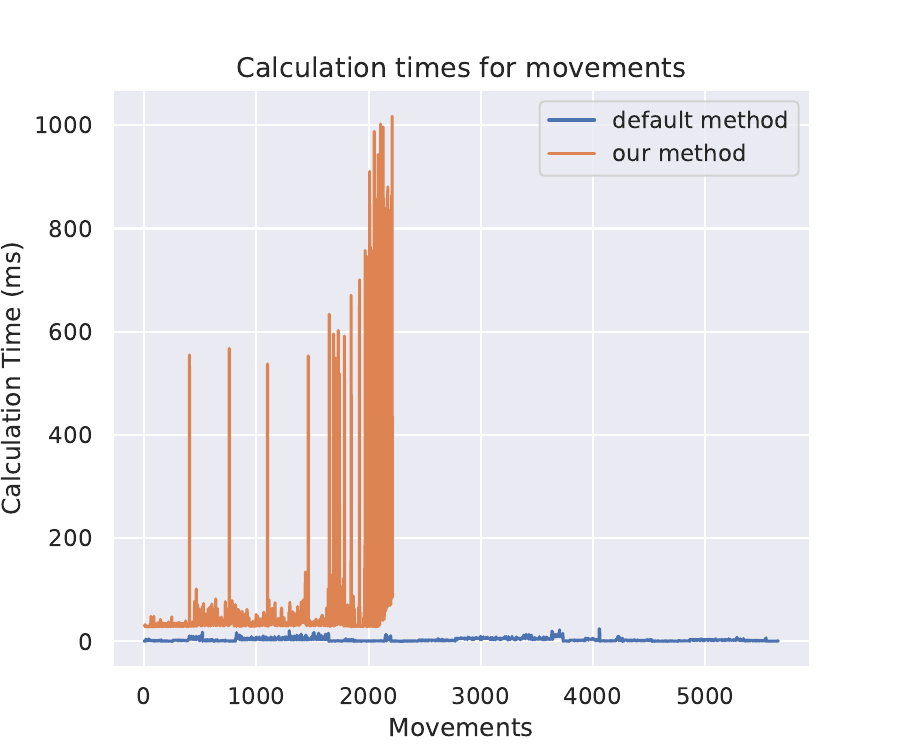}}
    \vspace{-1em}
    \caption{\label{fig:calctime} Calculation time required for each move in (a) cluster A and (b) cluster B. Our method requires more calculation time near termination, since more source devices are tried until the algorithm gives up.}
\end{figure*}

Movement calculation times for cluster A and cluster B are shown in Figure~\ref{fig:calctime}, (left) and (right), respectively.
For both clusters, the default balancer suggested movements faster with calculation times less than 2 ms for cluster A and 50 ms for cluster B, respectively.
For \textit{Equilibrium}, the higher one configures the $k$ parameter to try the $k$-fullest OSDs as candidates, the longer a movement generation can take before termination.
\textit{Equilibrium} required up to 10 ms for generating movement suggestions on cluster A and 1000 ms on cluster B, respectively.

\section{Discussion}
With ever increasing storage requirements on data centers, cost efficient utilization of storage devices is crucial.
Current data distribution algorithms for Ceph address object storage device (OSD) utilization insufficiently, resulting in less available storage and thus higher costs.
This paper presents \textit{Equilibrium}, a balancing algorithm for efficient object storage utilization.

While recent work aimed for optimizing several aspects of data distribution and access~\cite{flores2023readbalance,li2023adaptive,park2022reinforcement,wang2020mapx,wu2017boss,wu2017optimization}, a research gap remained on optimizing the available storage capacity by improving shard balancing compared to the default \texttt{mgr balancer}.
Our work fills this gap.

We conducted comprehensive experiments with six distinct clusters to assess the effect of our proposed balancer \textit{Equilibrium}, see Table~\ref{tab:moveresults}.
For all clusters, except cluster B, \textit{Equilibrium} increased the available storage more while requiring similar or less data movement.
In cluster B, many of the pools have 16 or less PGs, which is not enough to distribute the pool across the available 995 OSDs.
When the many-PG pools are distributed better, devices that had shards of few-PG pools will fill up more, resulting in less space for few-PG pools, but much increased space for the many-PG pools.
Since those few-PG pools can only allocate a few devices, options for moving them to an as-empty alternative OSD are limited.
In practice this does not matter much since the few-PG pools do not demand much space, and if they do, their PG count will be increased.

The OSD variances converging to zero after moving on cluster A (Figure~\ref{fig:clusterA}) and B (Figure~\ref{fig:clusterB}) according to the \textit{Equilibrium} balancer suggests that the calculated movements lead to a near equal distribution.
However, these storage improvements come at the cost of higher calculation times as indicated by the results shown in Figure~\ref{fig:calctime}.
As storage movements of several terabytes require more time than planning their movement with \textit{Equilibrium}, we argue that the increased planning time investment is negligible.

Overall, our results show that improved movement planning and load balancing increases available storage size, making existing clusters more effective and reducing the per byte storage cost.

Our study has the following limitations:
First, this study relied on simulating the resulting movement's effects, due to time considerations.
Second, more diverse clusters are necessary to test the balancer's robustness.
Finally, our proposed method occasionally required longer movement planning times.
These can be generally reduced by a more optimized implementation.

In conclusion, this paper presented \textit{Equilibrium}, a balancing algorithm for Ceph.
It optimizes not only for shard PG counts, but also minimizes the utilization variance.
Employing \textit{Equilibrium} on a variety of different types of clusters resulted in strong improvements in the amount of available space and the variance of utilization, leading to more available storage space.


\section*{Availability}
The code associated with this work is publicly available at \url{https://github.com/TheJJ/ceph-balancer}.

\bibliographystyle{plain}
\bibliography{literature.bib}

\end{document}